\documentclass[a4paper,10pt]{article}

\usepackage[utf8]{inputenc}
\usepackage {graphicx}
\usepackage {amsfonts}
\usepackage{amsmath}
\usepackage{url}
\usepackage{hyperref}
\usepackage{amsthm}
\usepackage{mathtools}
\usepackage{bbm}
\usepackage{dsfont}
\usepackage{textcomp}
\usepackage[colorinlistoftodos]{todonotes}
\usepackage{algorithm,algorithmic}
\usepackage{booktabs}
\usepackage{subfig}
\usepackage[font={footnotesize}]{caption}

\graphicspath{{./pics/}}

\newcommand{\numberset}{\mathbb}

\newcommand{\R}{\numberset{R}}

\newtheorem{theorem}{Theorem}[section]

\newcommand{\footremember}[2]{%
    \footnote{#2}
    \newcounter{#1}
    \setcounter{#1}{\value{footnote}}%
}

\setuptodonotes{fancyline, color=blue!30, size=\tiny} 

\title{Silkswap: An asymmetric automated market maker model for stablecoins}

\author{ Nicola Cantarutti\footremember{nicola}{Corresponding author email: nicolacantarutti@gmail.com},
  Alex Harker, 
  Carter Woetzel \\   
  \url{shadeprotocol.io}
  }

\begin{document}

\maketitle
 
\begin{abstract}
Silkswap is an automated market maker model designed for efficient stablecoin trading with minimal price impact. 
The original purpose of Silkswap is to facilitate the trading of fiat pegged stablecoins with the stablecoin Silk,
but it can be applied to any pair of stablecoins.
The Silkswap invariant is a hybrid function that generates an asymmetric price impact curve.
We present the derivation of the Silkswap model and its mathematical properties. We also compare different numerical methods used to solve the invariant equation. 
Finally, we compare our model with the well known Curve Finance model. 
\end{abstract} 

\vspace{0.4em}


\section{Introduction}

Decentralized EXchanges (DEXs) are currently the most popular application of Decentralized Finance (DeFi). 
Unlike centralized exchanges, DEXs are not based on a single centralized entity acting as custodians or intermediaries. 
Instead, on DEXs traders retain full control of their funds and private keys, 
and smart contracts execute trades for users in a neutral and automated fashion.
Most DEXs use Automated Market Maker (AMM) models to define the rules of trading, rather than 
relying on the order book model. 
Liquidity providers can deposit their tokens into liquidity pools in exchange for rewards coming from swap fees and token farming.
The AMM algorithmically computes the price of the tokens inside a liquidity pool only based on the token balance. 
A technical introduction on the mechanics of AMMs can be found in \cite{Angeris} and \cite{Mohan}.

One of the first AMMs to appear in DeFi is Uniswap \cite{Uni1}, that together with its upgraded version, Uniswap v2 \cite{Uni2}, is based on the 
Constant Product Market Maker (CPMM) model. 
This model assumes that the product of the quantities of tokens in a liquidity pool is constant,
which guarantees infinite liquidity inside the pool. CPMM works quite well for volatile tokens, as it promptly adapts the price in response to new trades. 
Nevertheless, it is not very suitable when trading low volatility or stable assets.

Another important AMM model is the Constant Sum Market Maker (CSMM), that assumes that the sum of the quantities of tokens inside the pool is constant.
The advantage of this model is that any trade has zero price impact, 
however it also has the inconvenience that it permits the complete drain of the entire liquidity inside the pool. 
For this reason, it is not used in practical applications. 

Low price impact is a good property when swapping between fiat pegged stablecoins. Curve Finance \cite{Curve1}, formerly Stableswap, 
and the new version Curve v2 \cite{Curve2} 
implement a Hybrid Function Market Maker (HFMM) model with the exact purpose of facilitating swaps between stablecoins.
This model combines CPMM and CSMM together in order to take advantage of both the infinite liquidity property of
CPMM and the zero price impact property of CSMM.
 
The main goal of the Silkswap model is to facilitate the trades of Silk with other fiat pegged stablecoins. 
However this model can be used to trade any pair of stablecoins.
In the current version, it is designed only for two-asset liquidity pools.
Silk \cite{Silk} is a privacy-preserving overcollateralized stablecoin developed by Shade Protocol and native to the Secret Network blockchain \cite{Secret}. 
The main feature of Silk is that it is a stablecoin pegged to a basket of global currencies and commodities using decentralized price feeds, 
creating a digital currency that serves as a hedge against single currency volatility.
We developed Silkswap as an HFMM model inspired by the Curve Finance v2 model. Additionally we introduced more flexibility in the shape of the invariant function, 
allowing for an asymmetric price impact curve. 
In this way, it is possible to discourage possible imbalances within the liquidity pool.\\
In the next sections we derive the Silkswap invariant and show its mathematical properties. We present numerical results for 
three different zero-finder algorithms and finally we compare our model with the Curve finance model.

\section{Silkswap model}

Let us consider a liquidity pool containing only two stablecoins. 
The two tokens in focus are token $X$, that represents any fiat pegged stablecoin, and token $Y$ that represents Silk. 
We use lower-case letters $x$ and $y$ to indicate the quantities of $X$ and $Y$.
Without loss of generality, let us quote the price of $X$ and $Y$ in US dollars, although any other fiat currency would work.
Let us introduce the conversion factors $p_X$ and $p_Y$, such that 
the values in dollars of $x$ and $y$ are simply $p_X \times x$ and $p_Y \times y$ respectively. 
The conversion factor $p_X$ has units [USD]/[X] and $p_Y$ has units [USD]/[Y].
We also introduce $p := \frac{p_Y}{p_X}$, the conversion factor between $Y$ and $X$, which has units [X]/[Y].\\ 
The conversion factors $p_X$ and $p_Y$ represent the market prices of one unit of $X$ and $Y$ respectively, while $p$
is the price of one unit of $Y$ in terms of $X$.
These values must be provided by an oracle, which is an external source of information that is fed into the AMM smart contract with a certain
frequency.

\noindent
\textbf{Example:}\\
Let us consider a liquidity pool containing USDC (token $X$) and Silk (token $Y$). 
If the oracle price of Silk is 1.05\$ then $p_Y = 1.05 \, \text{USD/SILK}$, 
while for USDC the oracle price is exactly 1\$, 
resulting in a perfect peg, then we have $p_X = 1 \, \text{USD/USDC}$. The direct conversion between Silk and USDC is therefore $p = \frac{p_Y}{p_X} = 1.05 \, \text{USDC/SILK}$.
Later we will often identify $X$ with USDC and $Y$ with SILK to make the discussion clearer.\\

Let us define the \textbf{equilibrium point} in a liquidity pool as the point $(x,y)$ where the dollar value of $x$ equals the dollar value of $y$.
The equilibrium point satisfies the \textbf{equilibrium equation}
\begin{equation}\label{eq_usd}
 p_Y \, y \, = \, p_X \, x.
\end{equation}
\noindent
From now on, we will assume that $X$ is our numeraire, and we will express the value of $Y$ in terms of $X$. The equilibrium equation becomes 
\begin{equation}\label{eq_p}
  p \, y = x.
 \end{equation}
\noindent
This choice is due to the fact that it is more natural for the user to evaluate an asset in terms of a fiat pegged stablecoin. 
Because Silk is pegged to a basket, its value expressed in terms of any fiat currency is not stable, but it fluctuates with a very low volatility.

\subsection{Silkswap invariant}


\begin{figure}[t!]
\hspace{-5em}
\begin{minipage}[b]{0.5\textwidth}
 \centering
 \includegraphics[scale=0.31]{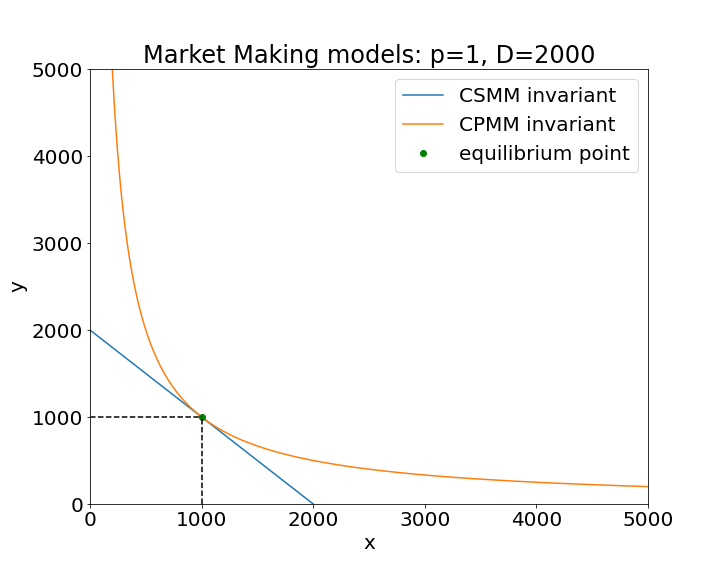}
 \end{minipage}
 \ \hspace{2.5em} \
 \begin{minipage}[b]{0.5\textwidth}
 \centering
   \includegraphics[scale=0.31]{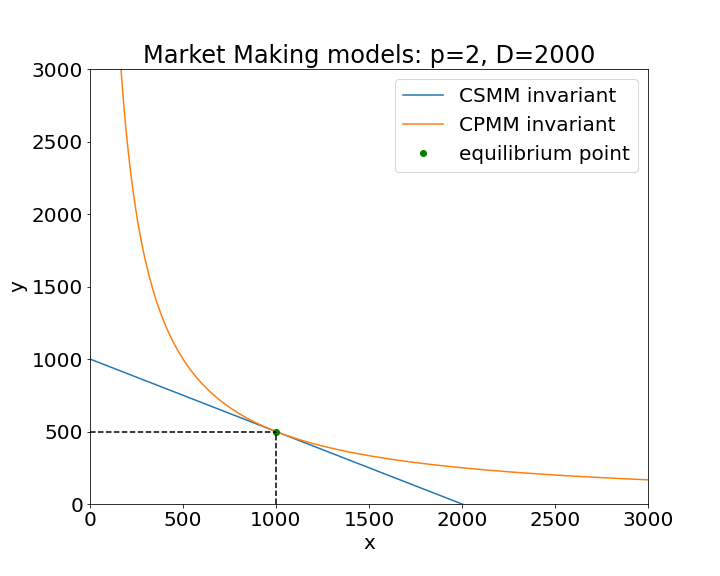}
 \end{minipage}
 \caption{Graph of the CPMM and CSMM models with $D=2000$. On the left $p = 1$. On the right $p = 2$.}
   \label{Fig1}
\end{figure}

Under the CPMM model, the quantities $x$ and $y$ must satisfy the following equation:
\begin{equation}\label{cpmm}
 x \, p \, y = K
\end{equation}
with $K>0$ constant. The equilibrium point, satisfying both (\ref{eq_p}) and (\ref{cpmm}), is $(x,y) = (\sqrt{K}, \frac{\sqrt{K}}{p})$. 
\noindent
Under the CSMM model instead, the following equation holds:
\begin{equation}\label{csmm}
  x \,+\, p y = D
 \end{equation}
with $D>0$ constant. The equilibrium point, satisfying (\ref{eq_p}) and (\ref{csmm}), is $(x,y) = (\frac{D}{2}, \frac{D}{2p})$.
The equilibrium point is a fundamental point where the ratio of the quantities of tokens in the pool is equal to $p$, and any invariant curve should contain this point. 
Since this point is unique, it follows that $K = \frac{D^2}{4}$.
In figure \ref{Fig1} we can see the graph of these two models, with different values of $p$.

\begin{figure}[t!]
\hspace{-5em}
\begin{minipage}[b]{0.5\textwidth}
 \centering
 \includegraphics[scale=0.31]{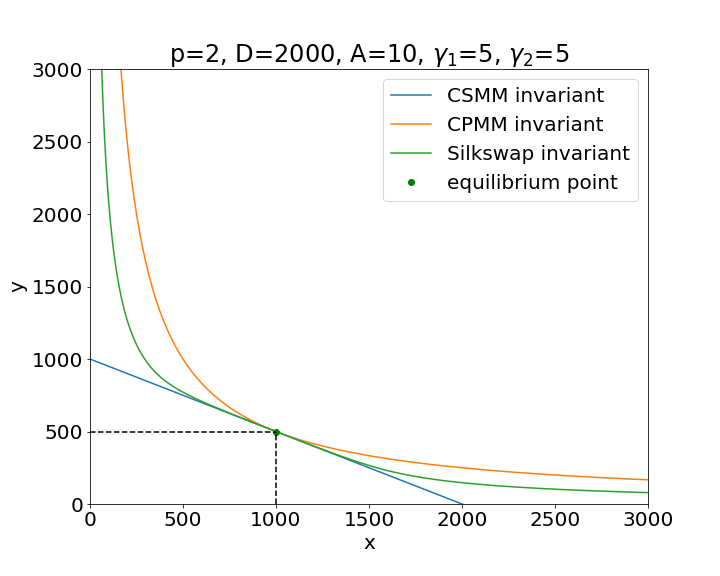}
 \end{minipage}
 \ \hspace{2.5em} \
 \begin{minipage}[b]{0.5\textwidth}
 \centering
   \includegraphics[scale=0.31]{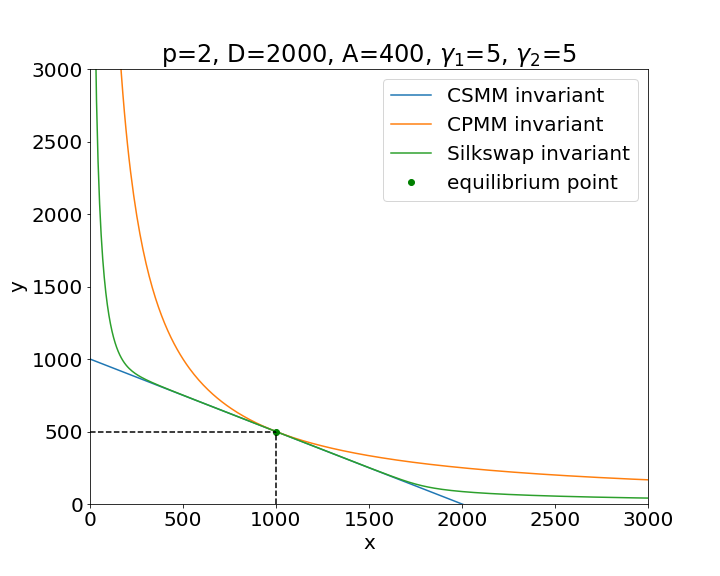}
 \end{minipage}
 \caption{Graph of the Silkswap invariant. On the $A=10$. On the right $A=400$. We can see that the parameter $A$ 
 indicates the closeness of the Silkswap curve to the CSMM line.}
   \label{Fig2}
 
 \hspace{-5em}
\begin{minipage}[b]{0.5\textwidth}
 \centering
 \includegraphics[scale=0.31]{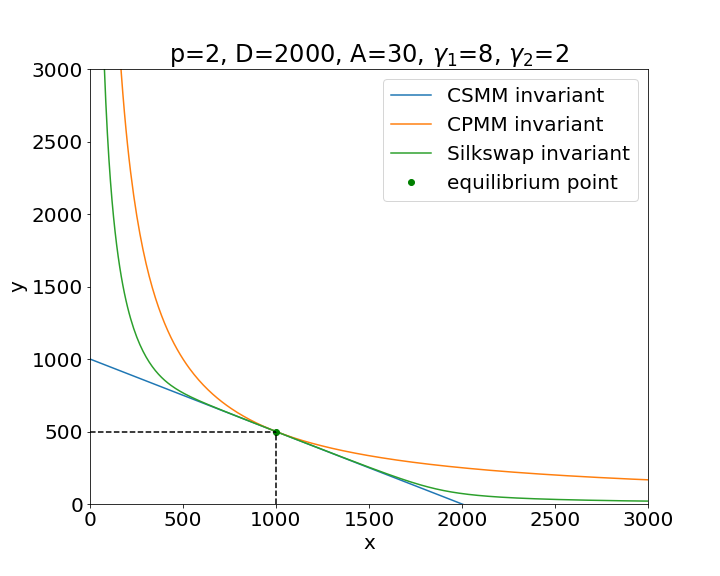}
 \end{minipage}
 \ \hspace{2.5em} \
 \begin{minipage}[b]{0.5\textwidth}
 \centering
   \includegraphics[scale=0.31]{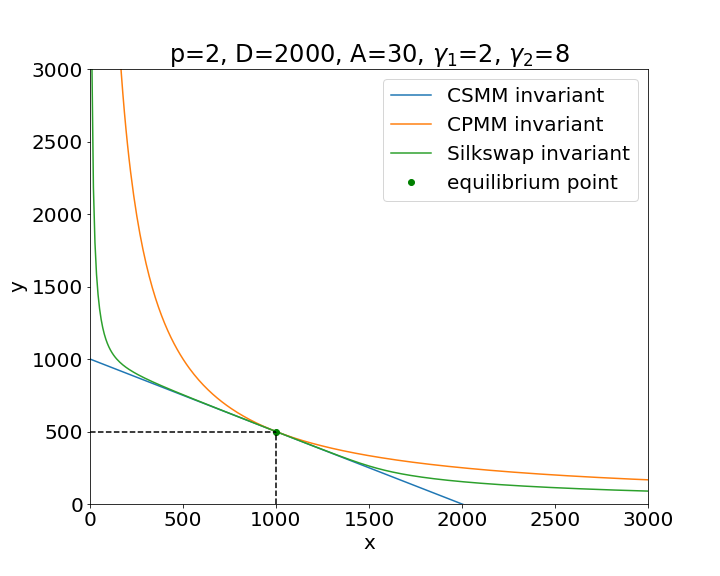}
 \end{minipage}
 \caption{Graph of the Silkswap invariant. We inverted the values of $\gamma_1$ and $\gamma_2$ to show how these parameters can control the asymmetry of the curve.}
   \label{Fig3}
\end{figure}

Now let us consider the following linear combination of the CSMM and CPMM models.
Let us multiply Eq. (\ref{csmm}) by $\chi A D$, and sum it with eq. (\ref{cpmm}):
\begin{equation}\label{stableswap}
  (\chi A D) \, (x + p y) + xpy \, = \, (\chi A D) D + \frac{D^2}{4}
 \end{equation}
where $\chi$ is a function of $x$ and $y$, defined as
\begin{equation}\label{chi}
  \chi := \biggl( \frac{4xpy}{D^2} \biggr)^{\gamma} 
\end{equation}
and
\begin{equation}\label{gamma}
 \gamma := \begin{cases} 
  \gamma_1, & \mbox{if } x \leq py \\ 
  \gamma_2, & \mbox{if } x > py 
  \end{cases} 
\end{equation} 
and the other parameters are constants satisfying $A>0$, $\gamma_1 \geq 0$, $\gamma_2 \geq 0$.
Let us remark that $\chi$ is an adimensional quantity and $D$ has the same dimension of $x$.\\
We call the equation (\ref{stableswap}), together with (\ref{chi}) and (\ref{gamma}), the \textbf{Silkswap invariant}. \\
The graph of the Silkswap invariant is shown in figures \ref{Fig2} and \ref{Fig3} under different set of parameters.
We can see that the CPMM hyperbola is always greater than the CSMM line, and they touch each other at the equilibrium point. The Silkswap invariant graph 
lies in between them. In the Appendix we prove these facts in the theorems (\ref{th_1}) and (\ref{th_3}). 
The parameter $A$ tells us how close the Silkswap graph is to the CPMM or to the CSMM.
From Eq. (\ref{stableswap}) we can easily see that for $A \to 0$ the invariant converges to the CPMM curve, while for $A \to \infty$ it converges to the CSMM curve.\footnote{The parameter 
$A$ has the same meaning of the parameter $A$ in the Stableswap model \cite{Curve1}. }\\
The two dimensional function $\chi$ in (\ref{chi}) is a discontinuous function. We show the surface plot in Fig. (\ref{Fig5}). 
In the Appendix we prove that, although the Silkswap invariant contains a discontinuous function, the points of our interest are regular points where the invariant is 
continuously differentiable.

\begin{figure}[t!]
\hspace{2em}
\begin{minipage}[b]{0.5\textwidth}
 \centering
 \includegraphics[scale=0.2]{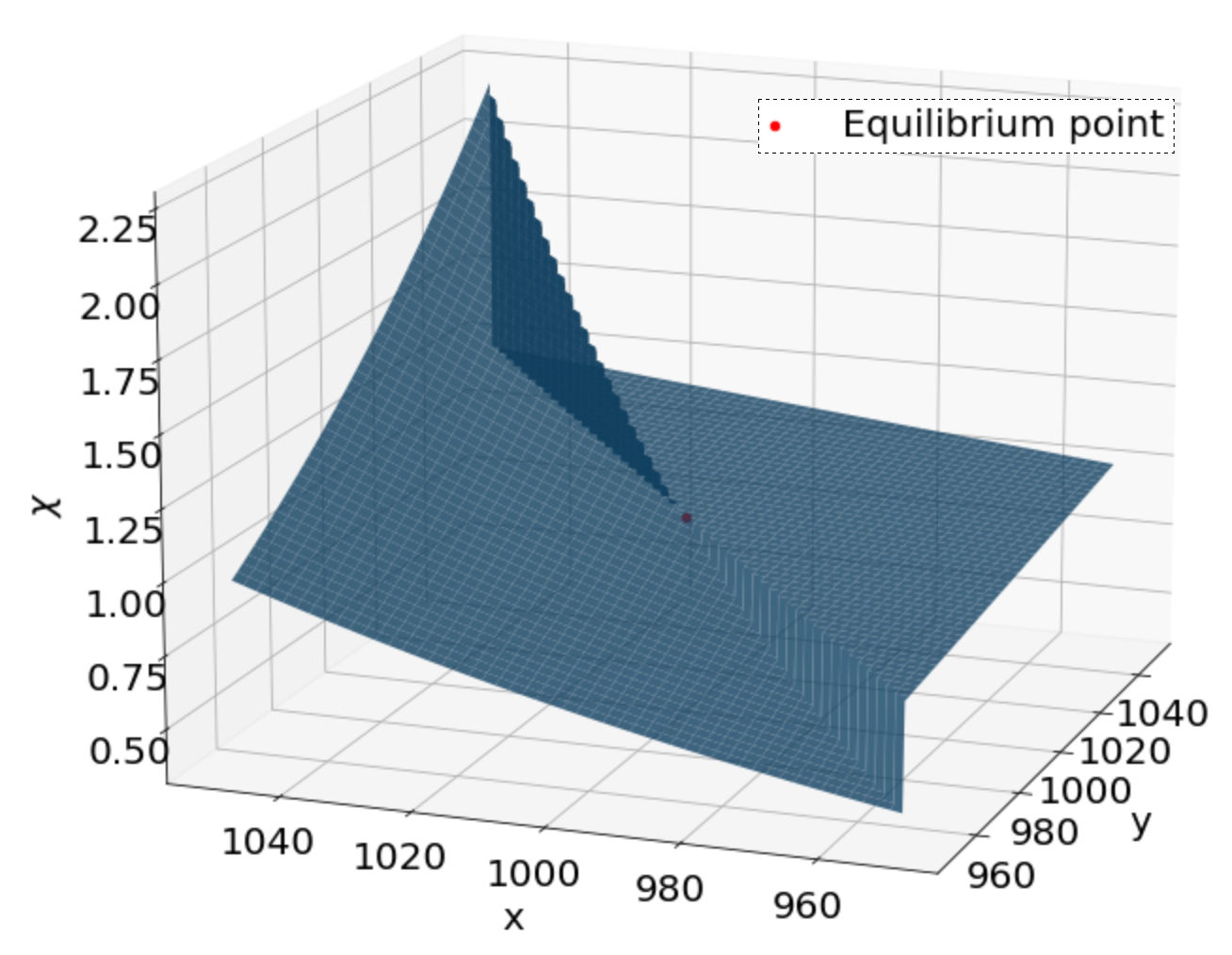}
\end{minipage}
 \caption{Function $\chi$ calculated with $D=2000$, $p=1$, $\gamma_1=2$, $\gamma_2=8$. The equilibrium point 
 is at $(x,y) = (1000,1000)$ and $\chi(x,y) = 1$. The function $\chi$ is continuous at this point, but not differentiable. }
   \label{Fig5}
\end{figure}

\subsection{Pricing with the Silkswap invariant}

In order to compute the price of a token, it is convenient to express the invariant in the explicit form $y = f(x)$, where $f : \R_{>0} \to \R_{>0}$ is 
continuously differentiable, decreasing and convex
\footnote{The set $\R_{>0}$ is the set $\{ x \in \R | x > 0 \}$.}. 
For the CPMM the explicit form is the hyperbola $f(x) = \frac{K}{p x}$, 
while for the CSMM the explicit form is the straight line $f(x) = \frac{1}{p}(-x + D)$, defined for $0 \leq x \leq D$.

If we trade an infinitesimal quantity $d y$, using the first order Taylor approximation $ d y \approx \frac{df(x)}{dx} d x$, it results equal to trade the quantity $\frac{df(x)}{dx} d x$.
Let us recall that by definition $\frac{df(x)}{dx}<0$, so we need to use the absolute value to define a positive price.
We define the \textbf{current price} of the token $X$ as function of $x$
\begin{equation}\label{price_X}
 P_X(x) \, := \, \biggl|\frac{df(x)}{dx}\biggr| \, = \, \biggl|\frac{dy}{dx}\biggr|.
\end{equation}
The price of the token $Y$ is defined as:
\begin{equation}\label{price_Y}
 P_Y(x)\, := \, \biggl|\frac{dx}{dy}\biggr| = \frac{1}{P_X}.
\end{equation}
Under the CPMM model, the current price is $P_Y(x) = \frac{p x^2}{K}$, and at equilibrium $P_Y(\sqrt{K}) = p$.
Under the CSMM model, the current price is a constant value, $P_Y(x) = p$, for any $0 < x < D$.\\

\begin{figure}[t!]
\hspace{-5em}
\begin{minipage}[b]{0.5\textwidth}
 \centering
 \includegraphics[scale=0.31]{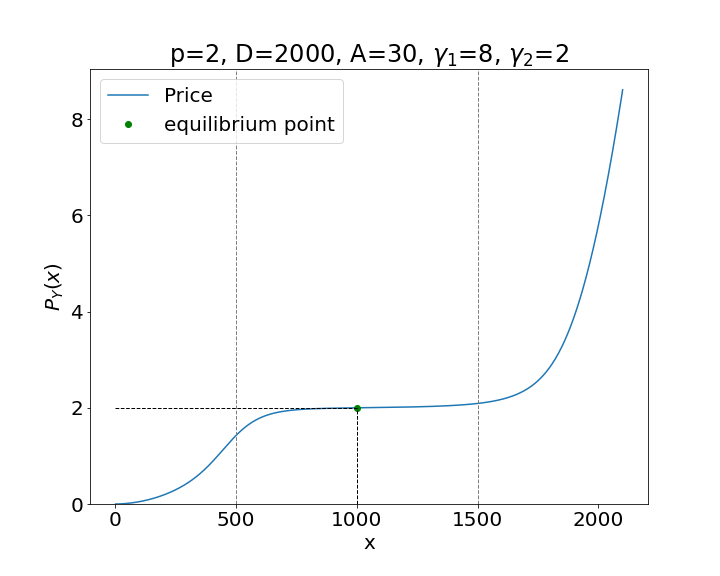}
 \end{minipage}
 \ \hspace{2.5em} \
 \begin{minipage}[b]{0.5\textwidth}
 \centering
   \includegraphics[scale=0.31]{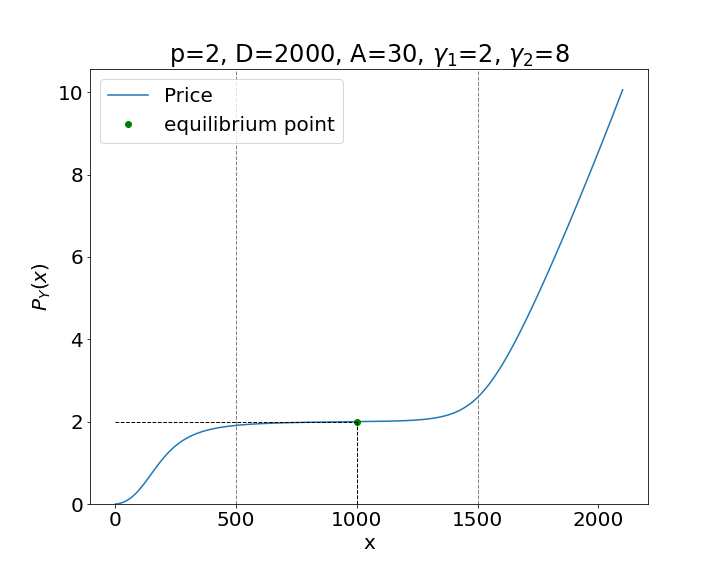}
 \end{minipage}
 \caption{Graph of the price of the token $Y$ as function of $x$, obtained from the Silkswap invariant. Same parameters used in Fig (\ref{Fig3}). The vertical lines at 
 500 and 1500 only serve to better understand the asymmetry in the graph.}
   \label{Fig4}
\end{figure}

Unfortunately the Silkswap invariant cannot be written in an explicit form. We can define the function $F: \R_{>0}^2 \to \R_{>0}$ as  
\begin{equation}\label{silkswap}
  F(x, y) \, := \, A D \, \biggl( \frac{4xpy}{D^2} \biggr)^{\gamma} \, (x + p y - D) + xpy  - \frac{D^2}{4}
\end{equation}
with $\gamma$ as in (\ref{gamma}). The Silkswap invariant (\ref{stableswap}) is the set of points that satisfy 
\begin{equation}\label{invariant_eq}
 F(x, y) = 0.
\end{equation}
The partial derivatives are:
\begin{align}\label{Fx} 
 \frac{\partial F}{\partial x} =& \; A D \biggl(\frac{4xpy}{D^2}\biggr)^{\gamma} \biggl[ 1 + \gamma \, 
				     \frac{x+p y-D}{x} \biggr] + py \\  
 \frac{\partial F}{\partial y} =& \; A D \biggl(\frac{4xpy}{D^2}\biggr)^{\gamma} \biggl[ p + \gamma \, 
				     \frac{x+p y-D}{y} \biggr] + px. \label{Fy}
\end{align}
\noindent
Thanks to the Theorem (\ref{C1}), we know the expression of the 
slope of the Silkswap invariant, which can be used to compute the prices of the tokens by (\ref{price_X}) and (\ref{price_Y}). The slope has the following
expression:
\begin{equation}\label{price_invariant}
 \frac{dy}{dx} = - \frac{ \frac{\partial F}{\partial x}  }{ \frac{\partial F}{\partial y} }\,.
\end{equation}
In Figure \ref{Fig4} we present two examples of price curves obtained from this expression.

\subsection{Scaled invariant}

\begin{figure}[t!]
\hspace{0em}
\begin{minipage}[b]{\textwidth}
 \centering
 \includegraphics[scale=0.3]{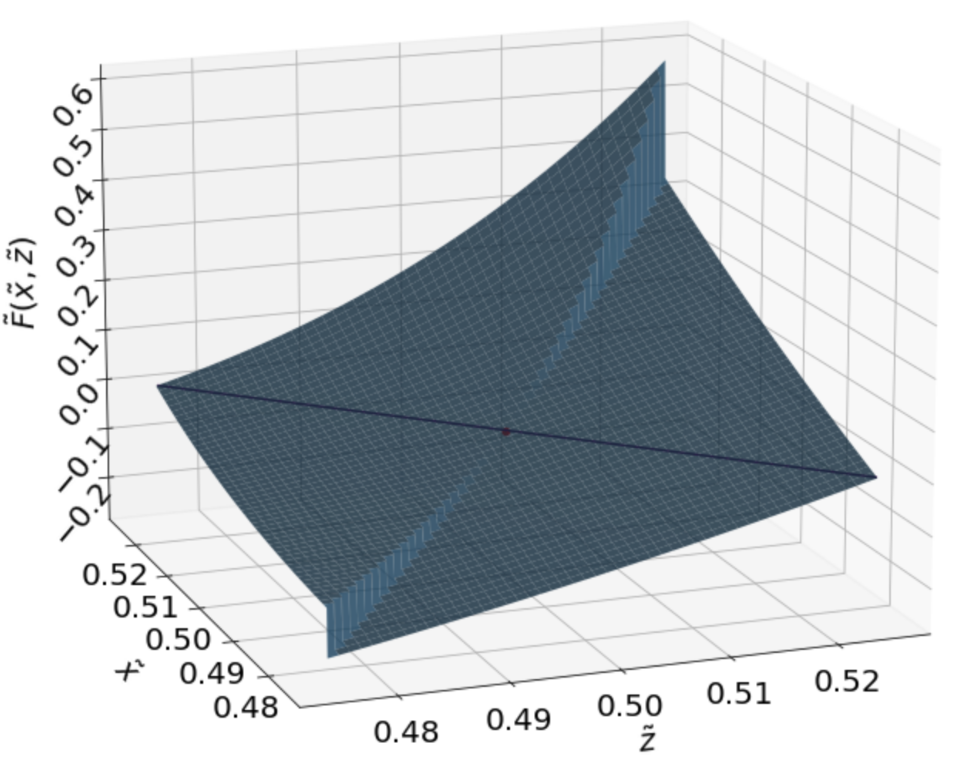}
\end{minipage}
 \caption{Scaled function $\tilde F(\tilde x, \tilde z)$, with, $\gamma_1=2$, $\gamma_2=8$. The equilibrium point 
 is at $(\tilde x,\tilde z) = (0.5,0.5)$ and the function is continuously differentiable in this point. The solid line represents the zero level of the invariant. }
   \label{Fig8}
\end{figure}

If we consider the function (\ref{silkswap}) depending also on $D$, for any constant $c > 0$ we have
\begin{equation}\label{scaling}
F(x, y, D) = 0  \quad \Longrightarrow \quad F(c x, c y, c D) = 0. 
\end{equation}
This means that the Silkswap invariant is also \textbf{invariant by scaling}.\\
This property is very useful in practice because it reduces the convergence time of some numerical methods 
and prevents possible overflows when the variables have very high magnitudes.\\ 
Let us introduce the variable $z := py$.
We can consider $F(c x, c y, c D)$ with $c = \frac{1}{D}$, equal to $F(\frac{x}{D}, \frac{y}{D}, 1)$, and define the corresponding scaled function:
\begin{equation}\label{scaled_fun}
\tilde F(\tilde x, \tilde z) \, := \, A \, \bigl( 4 \tilde x \tilde z \bigr)^{\gamma} \, (\tilde x + \tilde z - 1) + \tilde x \tilde z  - \frac{1}{4}.
\end{equation}
where $\tilde x = \frac{x}{D}$, and $\tilde z = \frac{z}{D}$.
We show in Fig. \ref{Fig8} the graph of this function.
The \textbf{scaled Silkswap invariant} is given by
\begin{equation}
 \tilde F(\tilde x, \tilde z) \, = \, 0.
\end{equation}
The function (\ref{scaled_fun}) is related with (\ref{silkswap}) by
\begin{equation}\label{homogeneous}
 \tilde F( \tilde x, \tilde z ) = \frac{1}{D^2} F(x, y). 
\end{equation}
Using the chain rule 
\begin{align}
& \frac{\partial \tilde F(\tilde x, \tilde z)}{\partial \tilde x} \, = \, \frac{\partial \tilde F(\tilde x, \tilde z)}{\partial x}\, \frac{d x}{d \tilde x} 
 \, = \, \frac{1}{D} \frac{\partial F(x, y)}{\partial x} \\
& \frac{\partial \tilde F(\tilde x, \tilde z)}{\partial \tilde z} \, = \, \frac{\partial \tilde F(\tilde x, \tilde z)}{\partial y}\, \frac{d y}{d \tilde z} 
 \, = \, \frac{1}{pD} \frac{\partial F(x, y)}{\partial y}, 
\end{align}
we can rewrite the price equation (\ref{price_invariant}) as
\begin{equation}
 \frac{dy}{dx} = - \frac{1}{p} \; \frac{ \frac{\partial \tilde F(\tilde x, \tilde z)}{\partial \tilde x}  }{ \frac{\partial \tilde F(\tilde x, \tilde z)}{\partial \tilde z} }\,.
\end{equation}
In the numerical calculation of the swap amount, we will make use of the scaled function (\ref{scaled_fun}), rather than (\ref{silkswap}), because it reduces a lot the number of 
operations, and consequently the run time and gas fees.

\section{Numerical implementation}

\subsection{Calculation of D}

\begin{figure}[t!]
\hspace{-5em}
\begin{minipage}[b]{0.5\textwidth}
 \centering
 \includegraphics[scale=0.31]{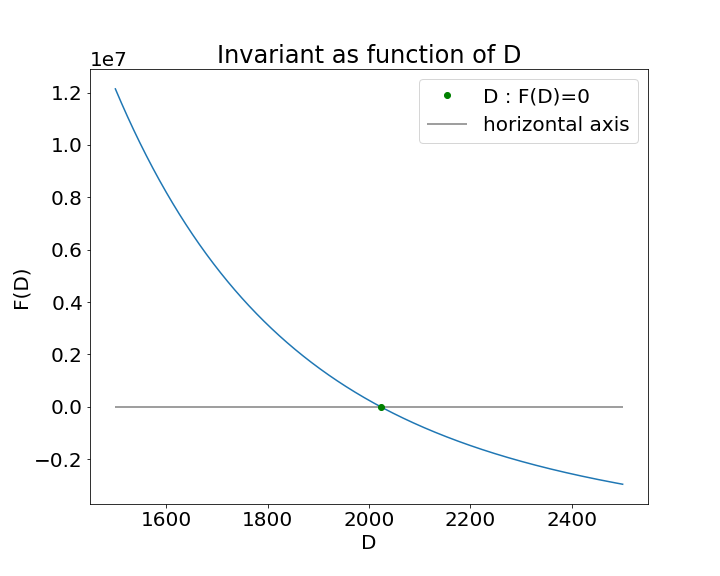}
 \end{minipage}
 \ \hspace{2.5em} \
 \begin{minipage}[b]{0.5\textwidth}
 \centering
   \includegraphics[scale=0.31]{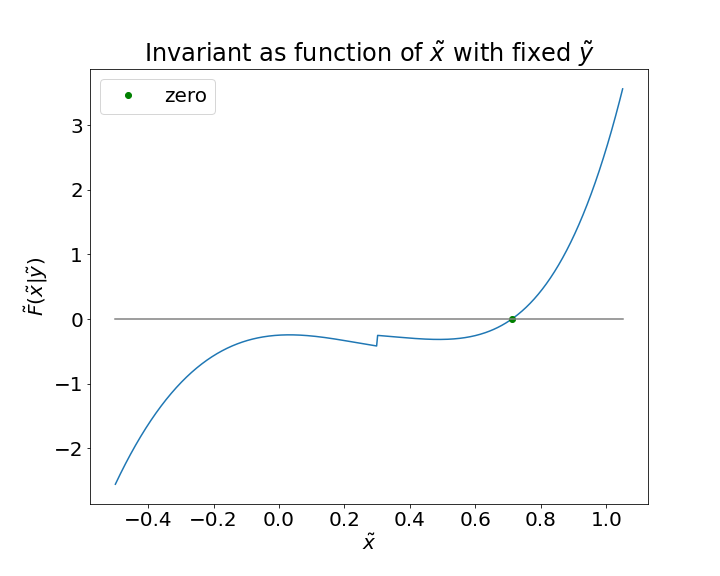}
 \end{minipage}
 \caption{LEFT: Invariant as function of $D$. We used $x=1900$, $y=100$, $p=2$, $A=30$, $\gamma_1=8$, $\gamma_2=2$. \\ 
 RIGHT: Scaled invariant as function of $\tilde x$. We used $y=600$, $D=2000$, $p=1$, $A=5$, $\gamma_1=2$, $\gamma_2=3$. The function has only one zero.}
\label{Fig6}

\hspace{-5em}
\begin{minipage}[b]{0.5\textwidth}
 \centering
 \includegraphics[scale=0.31]{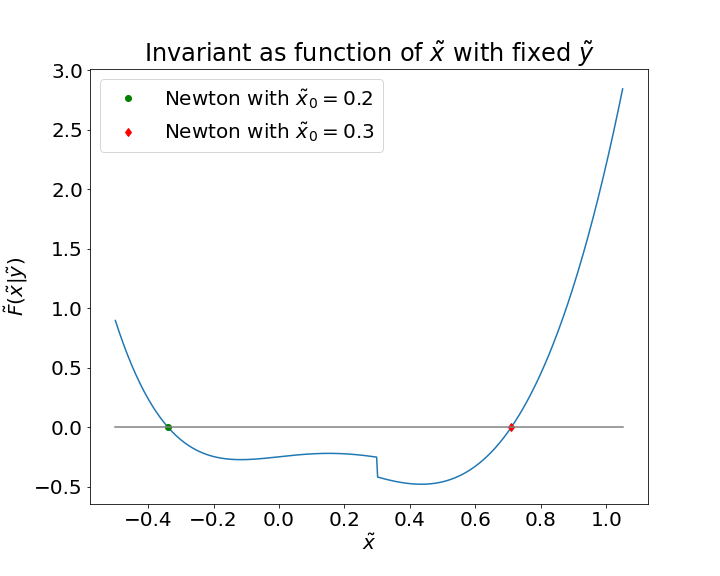}
 \end{minipage}
 \ \hspace{2.5em} \
 \begin{minipage}[b]{0.5\textwidth}
 \centering
   \includegraphics[scale=0.31]{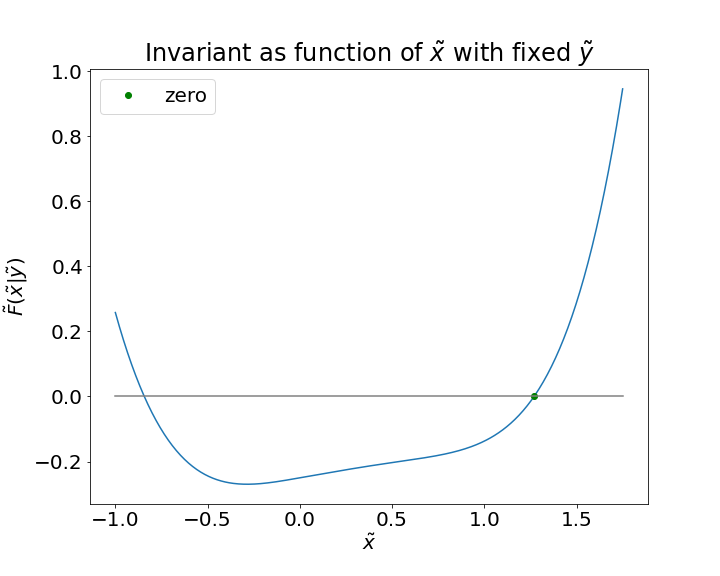}
 \end{minipage}
 \caption{ LEFT: Scaled invariant as function of $\tilde x$. We used $y=600$, $D=2000$, $p=1$, $A=5$, $\gamma_1=3$, $\gamma_2=2$. 
 The graph shows two zeros. 
 The Halley and Newton methods are highly dependent on the initial guess $\tilde x_0$.\\
 RIGHT: 
 Scaled invariant as function of $\tilde x$. We used $y=200$, $D=2000$, $p=1$, $A=5$, $\gamma_1=3$, $\gamma_2=4$. 
 The function is increasing for $\tilde x > 0$, therefore the Halley and Newton methods converge to the right solution. } 
\label{Fig7}
\end{figure}

Since this model is defined by an implicit equation, we need numerical methods to compute the variables of interest.
If the amounts of tokens in the pool satisfy the equilibrium equation \ref{eq_p}, then the parameter $D$ 
can be quickly computed from Eq. (\ref{csmm}). 
However, it is quite rare that a liquidity pool is in perfect equilibrium and the equation \ref{eq_p} is almost never satisfied.
In this cases we need to compute $D$ from Eq. (\ref{stableswap}) using a numerical method.
Let us define $F(D)$ as the function (\ref{silkswap}) when we consider $D$ variable and $x$, $y$ fixed. With an abuse of notation
we can call it "invariant function dependent of $D$", but let us recall that the name invariant can be used only when $F(D)=0$.
In Fig \ref{Fig6}, we can see that $F(D)$ is a smooth decreasing function.
We compare three different numerical methods: the Newton method, Halley method and the bisection method, see Table \ref{tab:D}. 
\begin{table}[ht]
\resizebox{0.5\linewidth}{!}{
\begin{tabular}[t]{cccc}
\toprule
\multicolumn{4}{c}{\textbf{Pool with 2000 USDC and 1000 SILK}} \\
\midrule
   Starting point    &  Method & Iterations &  Time \\
\midrule
2AM   & Newton & 4  & 312.8µs \\
2GM   & Newton & 8  & 770.4µs \\
2AM   & Halley & 2  & 390.5µs \\
2GM   & Halley & 5  & 820.4µs \\
 /    & Bisection & 61 & 2.37ms \\
\bottomrule
\end{tabular}
}
\quad
\resizebox{0.5\linewidth}{!}{
\begin{tabular}[t]{cccc}
\toprule
\multicolumn{4}{c}{\textbf{Pool with 200000 USDC and 100000 SILK}} \\
\midrule
   Starting point    &  Newton & Halley &  Bisection \\
\midrule
2AM   & Newton & 4 & 290.0µs \\
2GM   & Newton & 7 & 790.3µs \\
2AM   & Halley & 3 & 399.5µs \\
2GM   & Halley & 5 & 914.2µs \\
 /    & Bisection & 67 & 2.6ms
 \\
\bottomrule
\end{tabular}
}
\caption{Performance tables for the calculation of the parameter $D$. Newton and Halley methods are computed using two different initial guesses: $2\,AM$ and $2\,GM$. 
The bisection method is computed inside the interval $[$2GM, 2AM$]$, see Theorem \ref{D_bounds}. 
The parameters of the AMM are $p=1$, $A=100$, $\gamma_1 = \gamma_2 = 8$. We set a very small tolerance $\epsilon_D = 10^{-16}$. }
\label{tab:D}
\end{table}%

The idea of using two times arithmetic mean (AM) and geometric mean (GM) as starting points for Newton and Halley methods comes from Theorem \ref{D_bounds}. 
We can see that 2AM is a better choice. 
Although Newton method has more loop iterations than Halley, it performs better in terms of time. The reason is that the calculation of the second derivative is more expensive,
in terms of operations, than the extra iterations. The expression of the derivatives are in Appendix \ref{B}.
Bisection method, as expected from theoretical results, is the slowest. \\

\begin{table}[ht]
\resizebox{0.5\linewidth}{!}{
\begin{tabular}{*{11}l}
 \toprule
  \multicolumn{4}{c}{\textbf{Pool with $10^3$ USDC and $10^3$ SILK}} \\
  \midrule
  Swap size & Method &  Iterations & Time \\
  \midrule
    0         & Bisect & 0  & 153µs      \\  
    $0.1$     & Bisect & 52 & 1.32ms \\
    $1$       & Bisect & 52 & 1.31ms \\
    $10$      & Bisect & 52 & 1.30ms \\
    $10^2$    & Bisect & 52 & 1.37ms \\
    $10^3$    & Bisect & 52 & 1.31ms \\
    $10^4$    & Bisect & 52 & 1.34ms \\
    $10^5$    & Bisect & 52 & 1.47ms \\
    $10^6$    & Bisect & 52 & 1.46ms \\
    0         & Halley & 0  & 190µs  \\  
    $0.1$     & Halley & 2  & 311µs \\
    $1$       & Halley & 2  & 303µs \\
    $10$      & Halley & 3  & 366µs \\
    $10^2$    & Halley & 4  & 436µs \\
    $10^3$    & Halley & 9  & 751µs \\
    $10^4$    & Halley & 17 & 1.30ms \\
    $10^5$    & Halley & 26 & 1.95ms \\
    $10^6$    & Halley & 37 & 2.88ms \\
    0         & Newton & 0  & 159µs      \\  
    $0.1$     & Newton & 3  & 262µs \\
    $1$       & Newton & 3  & 246µs \\
    $10$      & Newton & 4  & 285µs \\
    $10^2$    & Newton & 6  & 358µs \\
    $10^3$    & Newton & 16 & 674µs \\
    $10^4$    & Newton & 30 & 1.20ms \\
    $10^5$    & Newton & 49 & 2.01ms \\
    $10^6$    & Newton & 68 & 2.80ms \\
    \bottomrule
  \end{tabular}
}
\quad
\resizebox{0.5\linewidth}{!}{
\begin{tabular}{*{11}l}
 \toprule
  \multicolumn{4}{c}{\textbf{Pool with $10^6$ USDC and $10^6$ SILK}} \\
  \midrule
  Swap size & Method &  Iterations & Time \\
  \midrule
    0         & Bisect & 0  & 141µs  \\  
    $0.1$     & Bisect & 52 & 1.51ms \\
    $1$       & Bisect & 52 & 1.45ms \\
    $10$      & Bisect & 52 & 1.62ms \\
    $10^2$    & Bisect & 52 & 1.46ms \\
    $10^3$    & Bisect & 52 & 1.48ms \\
    $10^4$    & Bisect & 52 & 1.46ms \\
    $10^5$    & Bisect & 52 & 1.57ms \\
    $10^6$    & Bisect & 52 & 1.47ms \\
    0         & Halley & 0  & 150µs  \\  
    $0.1$     & Halley & 1  & 270µs \\
    $1$       & Halley & 1  & 269µs \\
    $10$      & Halley & 2  & 346µs \\
    $10^2$    & Halley & 2  & 341µs \\
    $10^3$    & Halley & 2  & 341µs \\
    $10^4$    & Halley & 3  & 422µs \\
    $10^5$    & Halley & 4  & 488µs \\
    $10^6$    & Halley & 9  & 885µs \\
    0         & Newton & 0  & 155µs  \\  
    $0.1$     & Newton & 2  & 236µs \\
    $1$       & Newton & 2  & 239µs \\
    $10$      & Newton & 2  & 247µs \\
    $10^2$    & Newton & 3  & 297µs \\
    $10^3$    & Newton & 3  & 284µs \\
    $10^4$    & Newton & 4  & 316µs \\
    $10^5$    & Newton & 6  & 394µs \\
    $10^6$    & Newton & 16 & 1.04ms \\
  \bottomrule
  \end{tabular}
}
  \caption{Performance tables. We compare swap times where a trader swaps SILK for USDC. We consider different sizes and different numerical methods. 
  The initial guess for the Halley and Newton methods is $\tilde x_0$ i.e. the scaled amount of USDC before the swap. 
  The parameters of the AMM are $p=1$, $A=100$, $\gamma_1 = \gamma_2 = 8$. We set a very small tolerance $\epsilon_x = 10^{-16}$.}
  \label{tab:XY}
\end{table}

\subsection{Calculation of the swap amount}

We will also make use of numerical methods to calculate the amount of tokens that are returned during a swap.
When a quantity $\Delta y$ is introduced into the pool, we need to compute the quantity $\Delta x$ that is 
extracted from the pool, and such that the point $(x-\Delta x, y+\Delta y)$ belongs to the Silkswap invariant.
The parameter $D$ is fixed, and it is irrelevant for this calculation, therefore in order to simplify the problem, 
we can use the scaled function (\ref{scaled_fun}).
Let us define $\tilde F(\tilde x | \tilde y)$ and $\tilde F(\tilde y | \tilde x)$ the function with $\tilde y$ or $\tilde x$ fixed respectively. 
Again, with an abuse of notation
we can call these functions "invariant functions depending on $\tilde x$ or $\tilde y$".
In the following we consider the case when $\Delta x$ is extracted and therefore we need to find the zero of $\tilde F(\tilde x | \tilde y)$, but the same analysis works for 
$\tilde F(\tilde y | \tilde x)$.
In the Figures \ref{Fig6} and \ref{Fig7}, we can see that the function is discontinuous and can have different shapes depending on the choice of the parameters.
So the numerical convergence to the right solution is not always guaranteed. \\
By definition, the parameter $\gamma$ can be any non-negative real number. In practice we restrict $\gamma$ to be a positive integer value in order to simplify the model. 
The value of $\gamma_1$ is particularly important because it controls the behavior of the left tail: 
\begin{align*}
& \lim_{\tilde x \to -\infty} \tilde F(\tilde x | \tilde y) \to -\infty \quad \text{ for even } \gamma_1 \\ 
& \lim_{\tilde x \to -\infty} \tilde F(\tilde x | \tilde y) \to +\infty \quad \text{ for odd } \gamma_1, \\ 
\end{align*}
as we can see if we compare the plot of $\tilde F(\tilde x | \tilde y)$ in Fig. \ref{Fig6} with the plots in Fig. \ref{Fig7}.
When $\gamma_1$ is odd, $\tilde F(\tilde x | \tilde y)$ has two zeros. Since $\tilde x > 0$ by definition, we know that the correct solution must be positive. 
However, numerical methods such as Newton and Halley can converge to the wrong solution if initialized with an unlucky starting point, see Fig. \ref{Fig7} (LEFT).
Under some set of parameters, it is possible that the function $\tilde F(\tilde x | \tilde y)$ is increasing for $\tilde x > 0$, 
and the Newton method works fine, see Fig. \ref{Fig7} (RIGHT).
However, to avoid this uncertainty, in these cases it is better to use numerical methods that always guarantee the convergence to the right solution, such
as the bisection algorithm. If we call $(\tilde x_0, \tilde y_0)$ the amounts in the pool before the swap, 
and $(\tilde x_1, \tilde y_1)$ the amounts after the swap such that $\tilde F(\tilde x_1 | \tilde y_1)=0$,
then we know that $\tilde x_1 \in [0, \tilde x_0]$. The swap amount is therefore $\Delta x = D(\tilde x_0 - \tilde x_1)$. \\
In practical applications we want to take advantage of the speed of the Newton method, and therefore we choose to use only odd values for $\gamma_1$.
For comparisons between the three numerical methods under consideration, see Table \ref{tab:XY}.
Let us comment a few points:\\
\begin{enumerate}
 \item The number of iterations of the bisection method does not depend on the size of the pool. The reason is that we are searching in the interval $[0, \tilde x_0]$, and 
 $\tilde x_0 = \frac{x}{D}$. In the example, since we are at equilibrium $D=2x$ and therefore $\tilde x_0 = \frac{1}{2}$.
 \item The number of iterations of the bisection method does not depend on the size of the trade either.
 \item As for the calculation of $D$, Newton is slightly faster than Halley. The computation of second order derivatives (see Appendix \ref{B}) is expensive.
 \item The number of iterations of the Newton method increases when the size of the trade is big with respect to the size of the pool. In this case the Newton method is the one that
 performs worse while the bisection method is the one that performs best. 
 However, in practice it is very unlikely to see transactions bigger than the size of the pool.
 \item We used $\tilde x_0$ i.e. the value $\frac{x}{D}$ before the swap, as an initial guess. 
 We tried also with $\frac{\text{2AM}}{D}$ and $\frac{\text{2GM}}{D}$ as initial guesses, but the performances are worse.   
\end{enumerate}

When considering an even $\gamma_1$, it turns out that the Newton method is superior for both the calculations of $D$ and the swap size.  
In production, we decided to use Newton as main solver, and bisection as fallback in case of failure. 

\subsection{Model implementation}

We tested the model with values of $A$ ranging from 1 to $10^5$ and values of gamma from $1$ to $75$, 
under a variety of conditions including pool sizes from 1 to $10^{11}$ USDC total value, with a similar range of trade sizes. 
Generally, the Rust implementation can handle trade sizes several orders of magnitude above the size of the pool without overflowing, 
unless the values for $\gamma_1$ or $\gamma_2$ are excessively high. Since $\gamma_1$ and $\gamma_2$ are exponents, large values quickly create unmanageable numbers.

We implemented the Silkswap algorithm inside Rust smart contracts on the \textbf{Secret Network} blockchain. 
Smart contract do not allow floating point calculations, 
and this required the use of a few workarounds. 
Since smart contracts only support integer arithmetics, we stored most numbers as $10^{18}$ larger than their actual value.  
This effectively allowed us to store 18 decimal places. Variables are stored as \textbf{uint256} i.e. unsigned 256 bits integers.
Additionally, we paired each variable with a boolean representing its sign, allowing us to calculate both positive and negative numbers.
The square root in 2GM is computed by the Babilonian method.


\begin{figure}[t!]
\hspace{-5em}
\begin{minipage}[b]{0.5\textwidth}
 \centering
 \includegraphics[scale=0.31]{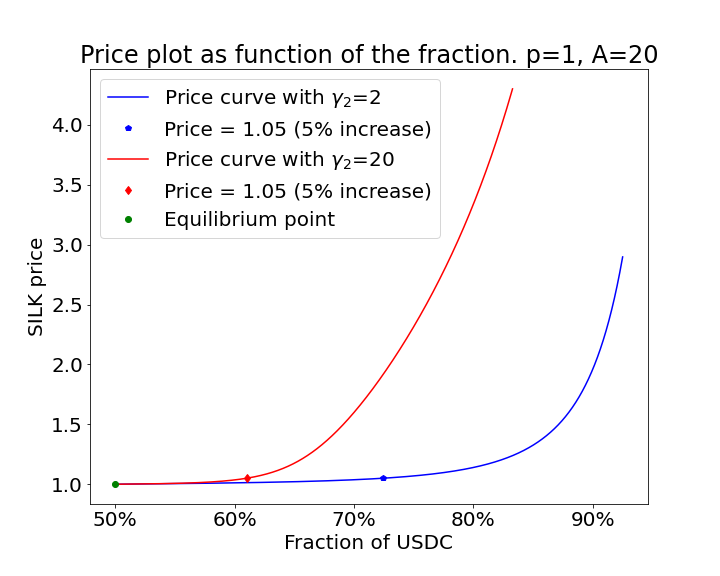}
 \end{minipage}
 \ \hspace{2.5em} \
 \begin{minipage}[b]{0.5\textwidth}
 \centering
   \includegraphics[scale=0.31]{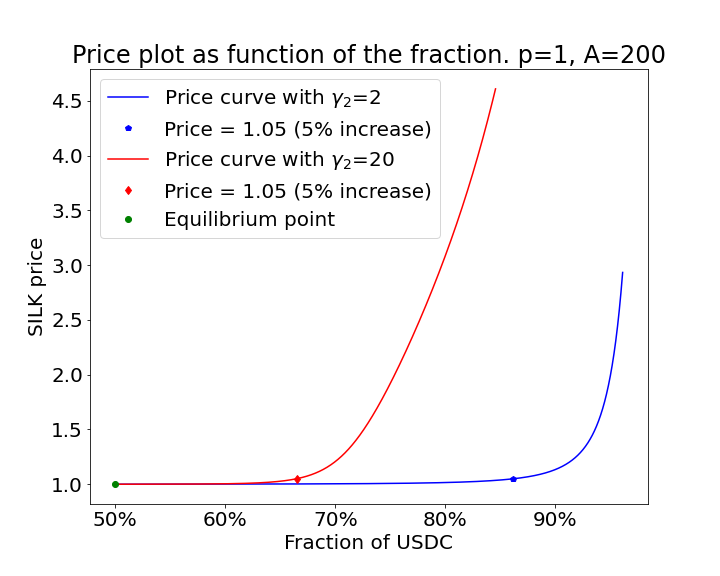}
 \end{minipage}
 \caption{Graph of the price of the token $Y$ (SILK) as function of the fraction of toke $X$ (USDC). Two lines corresponding to two different values of $\gamma_2$ are compared
 in order to show how this parameter can control the shape of the curve.}
   \label{Fig10}
\end{figure}

\subsection{Parameter selection}

In the previous sections we have seen how the parameters $A$, $\gamma_1$ and $\gamma_2$ modify the shape of the Silkswap invariant.
Specifically, parameter $A$ controls the flatness of the curve in the area around the equilibrium point (see Fig \ref{Fig2}), 
while the parameters $\gamma_1$ and $\gamma_2$ are used to control the curvature and asymmetry of the invariant (see Fig \ref{Fig3}). 
The shape of the invariant is directly reflected in the price curve (see Fig. \ref{Fig4}). 
Thus, the parameters of the model must be chosen in such a way as to make the price curve attractive for DEX users.
Therefore, they must maintain a low price impact and protect against possible imbalances in the liquidity pool.

In this section we present a practical method for deciding the value of parameters.
Since it is not very practical to consider a curve that depends on the quantity of tokens in the liquidity pool, 
we define the fraction of token $X$ inside the pool as $\frac{x}{x+py}$ and similarly we define the fraction of token $Y$ as 
$\frac{py}{x+py}$. As usual we multiply $y$ by $p$ to have the same units as $x$.
It is very convenient to express the price curve as a function of the fraction, because the fraction is a percentage and does not depend on the size of the pool.
Now we can choose a level of price impact that suits us, and choose the parameters in such a way that the price is impacted by this price impact 
at a certain level of fraction of the tokens.\\
In Fig. \ref{Fig10}, we show examples of price curves for different parameter values. We choose a 5\% level of price impact for $Y$.  
We are considering the case $x > py$ such that we are able to calculate $\gamma_2$. The same procedure should also be done for $x < py$ and considering a price impact 
for $X$ in order to find $\gamma_1$.\\
In this example, the price of the token $Y$ (SILK) at equilibrium is $p=1$. Its value after the 5\% price impact increase is $1.05$ and it is represented by the points 
on the price curves in the figure.
We can see how by changing $A$ and $\gamma_2$ it is possible to move the point 
to the right or to the left as desired. \\
A look at the curvature can also be helpful in choosing these parameters.

\section{Comparison with the Curve model}

Curve finance is currently the most popular DEX for trading stablecoins. 
The Curve v2 model \cite{Curve2} is an HFMM model, conceptually not very different from Silkswap.  
They both take a linear combination of the CSMM and the CPMM models. 
Since in the development of Silkswap we took inspiration from Curve, we decided to use a similar notation.
The parameter $A$ in Curve has the same meaning as in Silkswap.
The main difference between these models is the definition of the function $\chi$:
\begin{align*}
 & \text{Silkswap : } \quad   \chi = \biggl( \frac{4xy}{D^2} \biggr)^{\gamma}  \quad \quad \quad  \gamma = \begin{cases} 
  \gamma_1, & \mbox{if } x \leq y \\ 
  \gamma_2, & \mbox{if } x > y 
  \end{cases} \\
 & \text{Curve v2 : } \quad  \chi = \frac{4xy}{D^2} \biggl( \frac{\gamma}{\gamma+1-\frac{4xy}{D^2}} \biggr)^{2}. 
\end{align*}
For simplicity we didn't include the conversion factor $p$.
In both models $0 \leq \chi \leq 1$. 
If for a moment we do not consider the fact that our gamma can assume two values, the gammas in the two models have different meanings.
For any point not at equilibrium, in Silkswap: $\lim_{\gamma \to 0} \chi = 1$ and $\lim_{\gamma \to \infty} \chi = 0$, while for 
Curve v2 we have the opposite: $\lim_{\gamma \to 0} \chi = 0$ and $\lim_{\gamma \to \infty} \chi = 1$.

\begin{figure}[t!]
\hspace{-5em}
\begin{minipage}[b]{0.5\textwidth}
 \centering
 \includegraphics[scale=0.31]{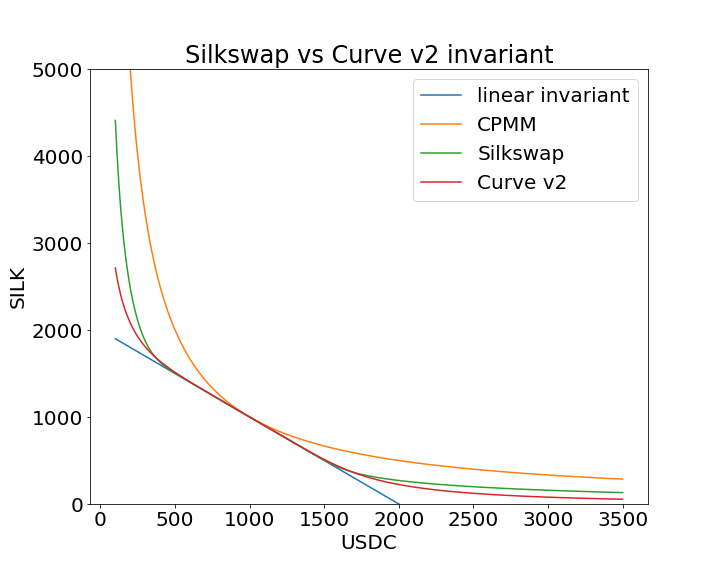}
 \end{minipage}
 \ \hspace{2.5em} \
 \begin{minipage}[b]{0.5\textwidth}
 \centering
   \includegraphics[scale=0.31]{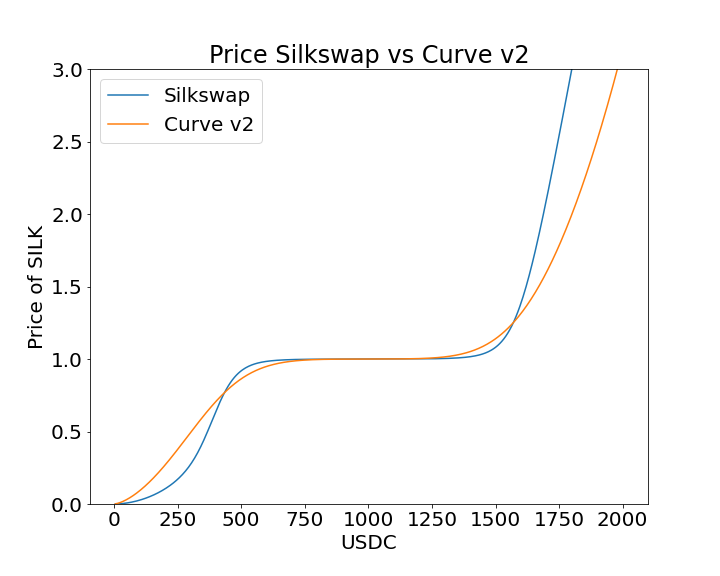}
 \end{minipage}
 \caption{LEFT: Comparison of the Silkswap invariant with the Curve v2 invariant curves. RIGHT: Comparison of price curves of SILK as function of USDC in the pool. 
 We used $D=2000$, $p=1$, and $A=400$ for both models. In Silkswap we use $\gamma_1 = \gamma_2 = 10$. In Curve v2 we use $\gamma=0.05$}
   \label{Fig9}
\end{figure}

In Fig. \ref{Fig9} we can see the invariant and price curves produced by these two models. The values of gammas are chosen to make the two curves as close as possible when using same 
value of $A$. 
We notice that the Silkswap model produces a higher curvature, although we did not prove it formally. 

Another difference between the models is that currently 
Silkswap is designed to work with liquidity pools containing only two tokens, while Curve v2 can have pools with any number of coins.  

Unlike Curve, we have decided to apply transaction fees to the token that is inserted into the pool, following the same approach used by Uniswap and Osmosis.

\section{Conclusions}

The Silkswap invariant is an AMM model that allows users to trade stablecoins with minimal price impact when the state of the pool is close to equilibrium. 
Away from equilibrium, the price impact tends to that of the CPMM model.
The main innovation of this model is the asymmetry of the invariant, which helps to regulate the liquidity in the pool and discourage strong imbalances in the quantity of tokens 
asymmetrically.
For new stablecoins like Silk that are in the process of building up liquidity, 
the asymmetric curve provides attack protection, discouraging the sale when the amount of Silk in the pool exceeds a certain threshold (dependent on the parameters of the model). 
This is done by making the price impact grow faster in that direction and slower in the opposite. Figure \ref{Fig4} shows this behavior very well.

\section*{Acknowledgements}
This research was fully funded by Shade Protocol.

\appendix
\numberwithin{equation}{section}

\section{Properties of the Silkswap invariant}

In Figure \ref{Fig1} we can see that the CPMM graph is greater than the CSMM graph, and they intersect each other at the equilibrium point. 
Let us formally prove this fact.
\begin{theorem}\label{th_1}
For $x > 0$, the condition 
\begin{equation}\label{cond1}
 \frac{1}{p}(-x + D) \; \leq \; \frac{D^2}{4px}
\end{equation}
is always satisfied.
\end{theorem}
\begin{proof}
The condition (\ref{cond1}) can be written as
$$ x^2 -Dx +\frac{D^2}{4} = \biggl( x - \frac{D}{2}\biggr)^2 \; \geq \, 0,$$
which is always verified.
\end{proof}

\begin{theorem}\label{AMGM}
The value of $\chi$, defined in (\ref{chi}) always satisfy
\begin{equation}\label{chi_0}
   0 < \chi \leq 1.
\end{equation}
\end{theorem}
\begin{proof}
Since $\chi$ is a function of only positive variables, it follows that $\chi$ must be positive.\\
Let us consider the Silkswap invariant (\ref{stableswap}):
\begin{equation}
 \underbrace{(\chi A D)}_{>0} \, \underbrace{(x + p y - D)}_{\geq_0} \,+\, \underbrace{xpy - \frac{D^2}{4} }_{\leq 0} \, = \, 0
\end{equation}
Since the sum of two terms is zero, it means that or both terms are zero, or the two terms have opposite sign. 
The case of both terms equal zero happens only at the equilibrium point. \\
Let us consider the case of both terms different than zero. We want to prove by contradiction that the first term must be positive.\\ 
If the first term is negative, then $(x + p y - D) < 0$, and we have that 
\begin{equation*}
 y \, <\, \frac{1}{p}(-x + D) \, \leq \, \frac{D^2}{4px},
\end{equation*}
by Theorem \ref{th_1}. This implies that $xpy - \frac{D^2}{4} < 0$, but this is a contradiction because the terms must have opposite sign.
The second term is always negative or zero, and we can conclude the proof. 
\begin{align*}
  xpy - \frac{D^2}{4} \, \leq \, 0 \quad \Longrightarrow \quad \frac{4xpy}{D^2} \; \leq \; 1 \quad \Longrightarrow \quad \chi \leq 1. 
\end{align*}
\end{proof}

In Figure \ref{Fig2} we can see that the Silkswap invariant always lies between the graphs of the CSMM and CPMM models. 
\begin{theorem}\label{th_3}
The graph of the Silkswap invariant always lies between the graphs of the CSMM and CPMM models.  
\end{theorem}
\begin{proof}
First we prove that the Silkswap invariant graph is not greater than the CPMM graph. This is a direct consequence of Theorem
\ref{AMGM}:
\begin{equation}\label{hyper_ineq}
 \frac{4xpy}{D^2} \leq 1 \quad \Longrightarrow \quad y \leq \frac{D^2}{4xp}.
\end{equation}
Now we prove that the Silkswap invariant graph is not smaller than the CSMM graph. Let us consider the invariant (\ref{stableswap}) and
divide it by the positive quantity $A D \chi$. We get 
\begin{align*}
 0 &= (x+py-D) + \frac{xpy - \frac{D^2}{4}}{AD\chi} \\
   & \leq x+py-D.
\end{align*}
where we used $xpy - \frac{D^2}{4} \leq 0$. It follows that 
\begin{equation}\label{linear_ineq}
 y \; \geq \; \frac{-x + D}{p}.
\end{equation}
\end{proof}

\begin{theorem}\label{D_bounds}
The parameter $D$ in the Silkswap invariant satisfies
\begin{equation}
 2\text{GM} \, \leq \, D \, \leq \, 2\text{AM},
\end{equation}
where GM is the geometric mean of $x$ and $py$, and AM is the arithmetic mean.
\end{theorem}
\begin{proof}
This is an immediate consequence of Theorem \ref{th_3}. From the two inequalities (\ref{hyper_ineq}) and (\ref{linear_ineq}) we 
obtain
\begin{equation}
 2 \sqrt{xpy} \, \leq \, D \, \leq \, x + py. 
\end{equation}
\end{proof}

\begin{theorem}\label{not_zero}
The partial derivatives (\ref{Fx}), (\ref{Fy}) of the function $F(x,y)$ defined in (\ref{silkswap}) are always positive on $\bigl\{ (x,y) \, :\, F(x,y)=0 \bigr\}$. 
\end{theorem}
\begin{proof}
We present a proof for $\frac{\partial F}{\partial y}$ only, since the same arguments can be used for $\frac{\partial F}{\partial x}$.
Let us consider the expression:
$$ \frac{\partial F}{\partial y} = \; A D \chi\, \biggl[ p (\gamma+1) 
				     +\gamma \biggl( \frac{x}{y} - \frac{D}{y}\biggr) \biggr] + px.$$
Since all variables are positive, when $\bigl( \frac{x}{y} - \frac{D}{y}\bigr) \geq 0$ then $\frac{\partial F}{\partial y} > 0$. This happens for $x \geq D$.\\
For $0<x<D$, let us rearrange the terms inside the square brackets and use (\ref{linear_ineq}):
\begin{align*}
 \biggl[ \gamma p + p - \gamma p \underbrace{ \biggl( \frac{1}{y} \, \frac{-x+D}{p} \biggr)}_{\leq 1} \biggr] \; \geq p \;> \;0.
\end{align*}
This last inequality proves the theorem.
\end{proof}   

Unfortunately, the function $F(x,y)$ defined in (\ref{silkswap}) is not continuous in the entire $\R_{>0}^2$, and we need to be careful
around the points of discontinuity. 
The following theorem guarantees the validity of (\ref{price_invariant}).

\begin{theorem}\label{C1}
The derivative of the Silkswap invariant can be written as
\begin{equation*}
 \frac{dy}{dx} = - \frac{ \frac{\partial F}{\partial x}  }{ \frac{\partial F}{\partial y} }\,.
\end{equation*}
\end{theorem}
\begin{proof}
The function $F(x,y)$ is continuously differentiable everywhere except on the line $x=py$, where it is discontinuous. 
The intersection between this line and the Silkswap invariant,
\begin{equation*}
\begin{cases} 
  x = py \\ 
  F(x, y) = 0, 
  \end{cases} 
\end{equation*} 
corresponds to the equilibrium point $(\frac{D}{2}, \frac{D}{2p})$. 
At the equilibrium point we have 
\begin{equation*}
 \lim_{ (x,y) \to \bigl(\frac{D}{2}, \frac{D}{2p}\bigr) } F(x,y)  \; = \; 0 
\end{equation*}
from any directions, and therefore $F(x,y)$ is continuous in this point. Also
\begin{align*}
 &\lim_{ (x,y) \to \bigl(\frac{D}{2}, \frac{D}{2p}\bigr) } \frac{\partial F}{\partial x} \; = \; \bigl(A + \frac{1}{2}\bigr) D \\
 &\lim_{ (x,y) \to \bigl(\frac{D}{2}, \frac{D}{2p}\bigr) } \frac{\partial F}{\partial y} \; = \; p \bigl(A + \frac{1}{2}\bigr) D
\end{align*}
with limits from any directions. Therefore the partial derivatives are continuous and $F(x,y)$ is continuously differentiable at $\bigl(\frac{D}{2}, \frac{D}{2p}\bigr)$.\\
Let us differentiate $F(x,y)$ along the direction of the Silkswap invariant
\begin{equation*}
 0 \;= \; d F(x,y) = \frac{\partial F}{\partial x} dx + \frac{\partial F}{\partial y} dy.
\end{equation*}
In the Theorem (\ref{not_zero}) we prove that $\frac{\partial F}{\partial y} > 0$. We can rearrange the last expression to conclude the proof.
\end{proof}   

\section{Expression of the derivatives}\label{B}

Derivative expressions used for the calculation of $D$ by Newton and Halley methods:

\begin{equation}\label{F1D}
 \frac{d F(D \,|\, x,y)}{d D} = A \biggl( \frac{4xpy}{D^2} \biggr)^{\gamma} \biggl[ (-2\gamma+1)\, (x + py- D) - D \biggr] - \frac{D}{2}, 
\end{equation}

\begin{equation}\label{F2D}
 \frac{d^2 F(D | x,y)}{d D^2} = A \biggl( \frac{4xpy}{D^2} \biggr)^{\gamma} \biggl[ 4\gamma-2 + 2\gamma(2\gamma-1) \biggl(\frac{x}{D} + \frac{py}{D} - 1\biggr)\biggr] - \frac{1}{2}.
\end{equation}

First derivative expressions for computing $\tilde x$ and $\tilde z$ by Newton and Halley methods: 

\begin{equation}\label{F1x}
 \frac{d \tilde F(\tilde x \,|\, \tilde z)}{d \tilde x} = A \bigl( 4\tilde x \tilde z \bigr)^{\gamma} \biggl[ \gamma\frac{\tilde x+\tilde z-1}{\tilde x} + 1\biggr] +\tilde z, 
\end{equation}

\begin{equation}\label{F1y}
 \frac{d \tilde F(\tilde z \,|\, \tilde x)}{d \tilde z} = A \bigl( 4\tilde x \tilde z \bigr)^{\gamma} \biggl[ \gamma\frac{\tilde x+\tilde z-1}{\tilde z} + 1\biggr] +\tilde x.
\end{equation}

for 
\begin{equation}\label{gammaa}
 \gamma := \begin{cases} 
  \gamma_1, & \mbox{if } \tilde x \leq \tilde z \\ 
  \gamma_2, & \mbox{if } \tilde x > \tilde z. 
  \end{cases} 
\end{equation} 
Second derivatives:

\begin{equation}\label{F2x}
 \frac{d^2 \tilde F(\tilde x \,|\, \tilde z)}{d \tilde x^2} = 4 \tilde z \gamma A \bigl( 4\tilde x \tilde z \bigr)^{\gamma-1} \biggl[ 2+ (\gamma-1) \frac{\tilde x+\tilde z-1}{\tilde x} \biggr],
\end{equation}

\begin{equation}\label{F2y}
 \frac{d^2 \tilde F(\tilde z \,|\, \tilde x)}{d \tilde z^2} = 4 \tilde x \gamma A \bigl( 4\tilde x \tilde z \bigr)^{\gamma-1} \biggl[ 2+ (\gamma-1) \frac{\tilde x+\tilde z-1}{\tilde z} \biggr].
\end{equation}


\bibliographystyle{apalike}
\bibliography{crypto.bib}

\begin{thebibliography}{}

\bibitem[Adams, 2018]{Uni1}
Adams, H. (2018).
\newblock Uniswap v1 whitepaper.
\newblock {\em \url{https://hackmd.io/@HaydenAdams/HJ9jLsfTz}}.

\bibitem[Adams et~al., 2020]{Uni2}
Adams, H., Zinsmeister, N., and Robinson, D. (2020).
\newblock Uniswap v2 core.
\newblock {\em \url{https://uniswap.org/whitepaper.pdf}}.

\bibitem[Angeris and Chitra, 2020]{Angeris}
Angeris, G. and Chitra, T. (2020).
\newblock Improved price oracles: Constant function market makers.
\newblock {\em Proceedings of the 2nd ACM Conference on Advances in Financial
  Technologies}, pages 80 -- 91.

\bibitem[Duniya, 2021]{Silk}
Duniya, S. (2021).
\newblock Silk: A privacy-preserving algorithmic burn stablecoin.
\newblock {\em \url{https://shadeprotocol.io/pdf/Silk_Whitepaper.pdf}}.

\bibitem[Egorov, 2019]{Curve1}
Egorov, M. (2019).
\newblock Stableswap - efficient mechanism for stablecoin liquidity.
\newblock {\em \url{https://curve.fi/files/stableswap-paper.pdf}}.

\bibitem[Egorov, 2021]{Curve2}
Egorov, M. (2021).
\newblock Automatic market-making with dynamic peg.
\newblock {\em \url{https://curve.fi/files/crypto-pools-paper.pdf}}.

\bibitem[Mohan, 2022]{Mohan}
Mohan, V. (2022).
\newblock Automated market makers and decentralized exchanges: a defi primer.
\newblock {\em Financial Innovation}, 8(20).

\bibitem[Woetzel, 2020]{Secret}
Woetzel, C. (2020).
\newblock Secret network: A privacy-preserving secret contract \& decentralized
  application platform.
\newblock {\em
  \url{https://www.securesecrets.org/Secret_Network_Graypaper_2.0.1_1.pdf}}.

\end{thebibliography}

\end{document}